\documentclass[a4paper,11pt,american]{article}
\usepackage{pos}

\usepackage{etoolbox}
\usepackage{ifthen}
\usepackage{xparse}

\usepackage{xspace}            
\xspaceaddexceptions{]}
\usepackage{xpunctuate}        
\usepackage[utf8]{inputenc}
\usepackage{babel}
\usepackage[T1]{fontenc}
\usepackage{csquotes}
\usepackage{lineno}

\usepackage{graphicx}  
\usepackage{xcolor}    
\usepackage[font=small]{caption}
\usepackage{subcaption}
\usepackage{booktabs}  
\usepackage{multirow}  

\usepackage{mathtools}          
\usepackage{amsfonts}           
\usepackage{amsxtra}            
\usepackage{xfrac}              
\usepackage{commath}            
\usepackage{braket}             

\usepackage[backend=biber,
    backref=false,
    style=numeric-comp,                                                                                                     sorting=none,
    block=ragged,
    giveninits=true,
]{biblatex}

\DeclareFieldFormat[article]{title}{\mkbibquote{#1}\isdot}
\DeclareFieldFormat[booklet]{title}{\mkbibquote{#1}\isdot}
\DeclareFieldFormat[report]{title}{\mkbibquote{#1}\isdot}
\DeclareFieldFormat[thesis]{title}{\mkbibquote{#1}\isdot}
\DeclareFieldFormat[misc]{title}{\mkbibquote{#1}\isdot}
\DeclareFieldFormat[book]{title}{\mkbibquote{#1}\isdot}
\DeclareFieldFormat[online]{title}{\mkbibquote{#1}\isdot}
\DeclareFieldFormat[phdthesis]{title}{\mkbibquote{#1}\isdot}
\DeclareFieldFormat[inproceeding]{title}{\mkbibquote{#1}\isdot}
\DeclareFieldFormat[incollection]{title}{\mkbibquote{#1}\isdot}
\DeclareFieldFormat[techreport]{title}{\mkbibquote{#1}\isdot}

\DeclareFieldFormat[article]{journaltitle}{#1\isdot}
\DeclareFieldFormat[article]{journalsubtitle}{#1\isdot}

\DeclareFieldFormat[article]{volume}{\textbf{#1}\isdot}

\renewbibmacro*{in:}{\printtext{\vphantom{a}}}

\renewbibmacro*{chapter+pages}{%
    \printfield{chapter}%
    \setunit{\addcomma\addspace}%
    \printfield{pages}%
    \newunit}

\DefineBibliographyStrings{UKenglish}{%
    page = {},
    pages = {}
}
\DefineBibliographyStrings{american}{%
    page = {},
    pages = {},
    backrefpage = {cited on page},
    backrefpages = {cited on pages}
}
\DefineBibliographyStrings{ngerman}{%
    page = {},
    pages = {}
}

\renewbibmacro*{institution+location+date}{
    \iflistundef{institution}
    {\setunit*{\addcomma\space}}
    {\setunit*{\addcolon\space}}%
    \printlist{institution}%
    \setunit*{\addcomma\space}%
    \usebibmacro{date}%
    \newunit}

\renewbibmacro*{pageref}{%
    \addperiod%
    \iflistundef{pageref}%
    {}%
    {\hfill\footnotesize\printtext[parens]{%
            \ifnumgreater{\value{pageref}}{1}%
            {\bibstring{backrefpages}\ppspace}%
            {\bibstring{backrefpage}\ppspace}%
            \printlist[pageref][-\value{listtotal}]{pageref}}\newunit}%
}

\begingroup
\global\let\savedifeof=\ifeof
\def\ifeof#1{\global\let\ifeof=\savedifeof\iftrue}%

\endgroup

\renewbibmacro*{author}{%
    \ifboolexpr{
        test \ifuseauthor
        and
        not test {\ifnameundef{author}}
    }
    {\printnames{author}%
        \iffieldundef{authortype}%
        {}%
        {\setunit{\printdelim{authortypedelim}}%
            \usebibmacro{authorstrg}}%
        \iffieldundef{collaboration}%
        {}%
        {\setunit{\printdelim{authortypedelim}}%
            \printtext[brackets]{\printfield{collaboration}}}%
    }
    {}}

\patchcmd{\bibsetup}{\interlinepenalty=5000}{\interlinepenalty=10000}{}{}  

\definecolor{MPPGreen}{RGB}{0,108,102} 
\definecolor{hyperrefLink}{RGB}{200,0,0}
\definecolor{hyperrefCite}{RGB}{0,170,0}
\definecolor{hyperrefUrl}{RGB}{0,170,170}

\usepackage[%
    textwidth=\marginparwidth,
    textsize=tiny,
    backgroundcolor=red!10,
    linecolor=red,
    bordercolor=red,
    english,
    obeyFinal]{todonotes}  

\usepackage{cleveref}

\usepackage{siunitx}
\DeclareSIUnit{\clight}{\text{\ensuremath{c}}}
\DeclareSIUnit[per-mode=symbol]\eVc{\eV\per\clight}
\DeclareSIUnit[per-mode=symbol]\meVc{\meV\per\clight}
\DeclareSIUnit[per-mode=symbol]\keVc{\keV\per\clight}
\DeclareSIUnit[per-mode=symbol]\MeVc{\MeV\per\clight}
\DeclareSIUnit[per-mode=symbol]\GeVc{\GeV\per\clight}
\DeclareSIUnit[per-mode=symbol]\TeVc{\TeV\per\clight}
\DeclareSIUnit[per-mode=symbol]\eVcc{\eV\per\clight\squared}
\DeclareSIUnit[per-mode=symbol]\meVcc{\meV\per\clight\squared}
\DeclareSIUnit[per-mode=symbol]\keVcc{\keV\per\clight\squared}
\DeclareSIUnit[per-mode=symbol]\MeVcc{\MeV\per\clight\squared}
\DeclareSIUnit[per-mode=symbol]\GeVcc{\GeV\per\clight\squared}
\DeclareSIUnit[per-mode=symbol]\TeVcc{\TeV\per\clight\squared}
\DeclareSIUnit[per-mode=symbol]\pereVcc {(\eV\per\clight\squared)^{-1}}
\DeclareSIUnit[per-mode=symbol]\permeVcc{(\meV\per\clight\squared)^{-1}}
\DeclareSIUnit[per-mode=symbol]\perkeVcc{(\keV\per\clight\squared)^{-1}}
\DeclareSIUnit[per-mode=symbol]\perMeVcc{(\MeV\per\clight\squared)^{-1}}
\DeclareSIUnit[per-mode=symbol]\perGeVcc{(\GeV\per\clight\squared)^{-1}}
\DeclareSIUnit[per-mode=symbol]\perTeVcc{(\TeV\per\clight\squared)^{-1}}
\DeclareSIUnit[per-mode=symbol]\pereVccsqrt {(\eV\per\clight\squared)^{-1/2}}
\DeclareSIUnit[per-mode=symbol]\permeVccsqrt{(\meV\per\clight\squared)^{-1/2}}
\DeclareSIUnit[per-mode=symbol]\perkeVccsqrt{(\keV\per\clight\squared)^{-1/2}}
\DeclareSIUnit[per-mode=symbol]\perMeVccsqrt{(\MeV\per\clight\squared)^{-1/2}}
\DeclareSIUnit[per-mode=symbol]\perGeVccsqrt{(\GeV\per\clight\squared)^{-1/2}}
\DeclareSIUnit[per-mode=symbol]\perTeVccsqrt{(\TeV\per\clight\squared)^{-1/2}}
\DeclareSIUnit[per-mode=symbol]\eVcsq{(eV/\clight)^2}
\DeclareSIUnit[per-mode=symbol]\meVcsq{(meV/\clight)^2}
\DeclareSIUnit[per-mode=symbol]\keVcsq{(keV/\clight)^2}
\DeclareSIUnit[per-mode=symbol]\MeVcsq{(MeV/\clight)^2}
\DeclareSIUnit[per-mode=symbol]\GeVcsq{(GeV/\clight)^2}
\DeclareSIUnit[per-mode=symbol]\TeVcsq{(TeV/\clight)^2}
\DeclareSIUnit[per-mode=symbol]\pereVcsq{(eV/\clight)^{-2}}
\DeclareSIUnit[per-mode=symbol]\permeVcsq{(meV/\clight)^{-2}}
\DeclareSIUnit[per-mode=symbol]\perkeVcsq{(keV/\clight)^{-2}}
\DeclareSIUnit[per-mode=symbol]\perMeVcsq{(MeV/\clight)^{-2}}
\DeclareSIUnit[per-mode=symbol]\perGeVcsq{(GeV/\clight)^{-2}}
\DeclareSIUnit[per-mode=symbol]\perTeVcsq{(TeV/\clight)^{-2}}
\DeclareSIUnit\mbarn{\milli\barn}
\DeclareSIUnit\mubarn{\micro\barn}
\DeclareSIUnit\nbarn{\nano\barn}
\DeclareSIUnit\pbarn{\pico\barn}
\DeclareSIUnit\fbarn{\femto\barn}
\DeclareSIUnit\abarn{\atto\barn}
\DeclareSIUnit[per-mode=reciprocal]\ipbarn{\per\pbarn}
\DeclareSIUnit[per-mode=reciprocal]\ifbarn{\per\fbarn}
\DeclareSIUnit[per-mode=reciprocal]\iabarn{\per\abarn}
\DeclareSIUnit\arbitaryunit{\text{a.u.}}
\DeclareSIUnit\counts{\text{units}}
\DeclareSIUnit\units{\text{units}}
\DeclareSIUnit\permil{\text{\textperthousand}} 
\DeclareSIUnit{\CPUh}{CPUh}

\usepackage{physics}

\usepackage[italic]{hepnames}
\RenewDocumentCommand{\Pai}{s o}{\IfBooleanTF{#1}{\HepParticle{a}{1}{}}{\HepParticleResonanceFull{a}{1}{}{\IfNoValueTF{#2}{1260}{#2}}{}{}}\xspace}
\RenewDocumentCommand{\Pfz}{s o}{\IfBooleanTF{#1}{\HepParticle{f}{0}{}}{\HepParticleResonanceFull{f}{0}{}{\IfNoValueTF{#2}{980}{#2}}{}{}}\xspace}
\RenewDocumentCommand{\PKzero}{s o}{\IfBooleanTF{#1}{\HepParticle{K}{}{}}{\HepParticleResonanceFull{K}{}{}{\IfNoValueTF{#2}{500}{#2}}{}{}}\xspace}
\RenewDocumentCommand{\PKstz}{s o}{\IfBooleanTF{#1}{\HepParticle{K}{0}{*}}{\HepParticleResonanceFull{K}{0}{*}{\IfNoValueTF{#2}{1430}{#2}}{}{}}\xspace}
\RenewDocumentCommand{\PKi}{s o}{\IfBooleanTF{#1}{\HepParticle{K}{1}{}}{\HepParticleResonanceFull{K}{1}{}{\IfNoValueTF{#2}{1270}{#2}}{}{}}\xspace}
\RenewDocumentCommand{\PKsti}{s o}{\IfBooleanTF{#1}{\HepParticle{K}{1}{*}}{\HepParticleResonanceFull{K}{1}{*}{\IfNoValueTF{#2}{892}{#2}}{}{}}\xspace}

\usepackage{ifdraft}
\newtoggle{debugging}
\makeatletter
\@ifclasswith{scrartcl}{draft}{%
	\toggletrue{debugging}%
}{%
	\@ifclasswith{scrartcl}{final}{%
		\togglefalse{debugging}%
	}{%
		\toggletrue{debugging}
	}%
}
\makeatother

\title{Studies of hadron spectroscopy at Belle and Belle~II}

\author*[a]{S. Wallner}
\onbehalf{on behalf of the Belle~II collaboration}

\affiliation[a]{Max Planck Institute for Physics,\\
  Boltzmanstr. 8, Garching, Germany}

\emailAdd{swallner@mpp.mpg.de}

\abstract{The Belle and Belle II experiments have collected a $1.6\,\mathrm{ab}^{-1}$ sample of $e^+e^-$ collision data at center-of-mass energies near the $\Upsilon(nS)$ resonances. We conduct searches for transitions from the spin-singlet $h_b(1P,2P)$ states to the spin-triplet $\Upsilon(1S)$ and $\chi_{bJ}(1P)$ states. We do not find evidence for the  $h_b(1P,2P)\to \Upsilon(1S) \pi^0$ and $h_b(2P) \to \chi_{bJ}(1P)\gamma$ transitions. We find the first evidence for the $h_b(2P)\to \Upsilon(1S)\eta$ transition. However, the measured branching fraction is lower than expected from related decays. Furthermore, we find evidence for $P_{c\bar c s}(4459)^0 \to J/\psi \Lambda$ decays in inclusive $\Upsilon(1S,2S)$ decays. This is the first evidence for an exotic state produced in $\Upsilon(1S,2S)$ decays.}

\FullConference{The European Physical Society Conference on High Energy Physics (EPS-HEP2025)\\
7-11 July 2025\\
Marseille, France\\}

\begin{document}
\maketitle

\section{The Belle and Belle~II Experiments}

The Belle experiment~\cite{Belle:2000cnh}, situated at the KEKB~\cite{Kurokawa:2001nw} accelerator facility in Tsukuba, Japan, collected a data sample of $e^+e^-$ collisions with center-of-mass energies near the $\Upsilon(nS)$ resonances corresponding to about \SI{1000}{\ifbarn} in the years 1999--2010.
In particular, it recorded the world's largest samples with center-of-mass energies at the $\Upsilon(1S)$, $\Upsilon(2S)$, and $\Upsilon(5S)$ resonance of \SI{6}{\ifbarn}, \SI{25}{\ifbarn}, and \SI{121}{\ifbarn}, respectively.
These large samples, in conjunction with Belle's excellent reconstruction of charged and neutral final-state particles, facilitate studies of transitions of hadrons and searches for exotic forms of hadrons that appear in the decays of these $\Upsilon(nS)$ resonances with unprecedented precision.

The Belle~II experiment~\cite{Belle-II:2010dht}, the successor to Belle, at the upgraded SuperKEKB~\cite{Akai:2018mbz} accelerator facility completed its first data taking run from 2019--2022 and collected already about half of the full Belle data set. The majority of this data was collected with a center-of-mass energy at the $\Upsilon(4S)$ resonance. Belle~II started its second run of data taking in the spring of 2024. At the end of 2024, the SuperKEKB accelerator achieved the current world record luminosity of \SI{5.1e34}{\per\cm\squared\per\second}. Belle~II will continue to collect data with the final goal of recording a data sample that is 50 times larger than the one previously recorded by Belle.

\section{Study of Spin Singlet to Triplet Transitions in $h_b(1P,2P)$ Decays}
\label{sec:bb}

In the quark model, bottomonium states are systems comprising a $b$ and anti-$b$ quark, bound by the strong interaction, and characterized by their quantum numbers. The spins of the quarks couple to a spin singlet ($S_{b\bar b} = 0$) or a spin triplet ($S_{b\bar b} = 1$) state.
Together with the orbital angular momentum between the $b$ and the anti-$b$ quark, it constitutes the total spin of the bottomonium.
A variety of bottomonium states with distinct quantum numbers exist. The study of transitions between these states enables the probing of their inner structure and the testing of effective field theories that are employed to predict the properties of these transitions in the non-perturbative regime of QCD.
Transitions between spin-singlet and spin-triplet states are of particular interest because they are suppressed by heavy-quark spin symmetry in the quark model. However, this suppression may be lifted by couple channel effects from hadron loops that can appear in these transitions~\cite{Guo:2016yxl}. A precise measurement of the branching fraction of such transitions will constrain these effects.

\subsection{Evidence for $h_b(2P)\to\Upsilon(1S)\eta$ and Search for $h_b(1P,2P)\to \Upsilon(1S) \pi^0$}
\label{sec:U1}

The decay of the spin-singlet state $h_b(2P)$ to the spin-triplet state $\Upsilon(1S)$ and an $\eta$ is an example of such a decay that is suppressed by heavy-quark spin symmetry.
However, the authors of ref.~\cite{Li:2012is} predict a large branching fraction of this decay of $\mathcal{B}[h_b(2P)\to \Upsilon(1S)\eta] \approx \SI{10}{\percent}$ based on evidence of the decay $\Upsilon(3S)\to h_b(1P)\pi^0$ observed by BaBar with a branching fraction of $\mathbf{B}[\Upsilon(3S)\to h_b(1P)\pi^0] \approx 10^{-3}$~\cite{BaBar:2011ljf}.

\begin{figure}
    \begin{subfigure}{0.49\linewidth}
    \includegraphics[width=\linewidth]{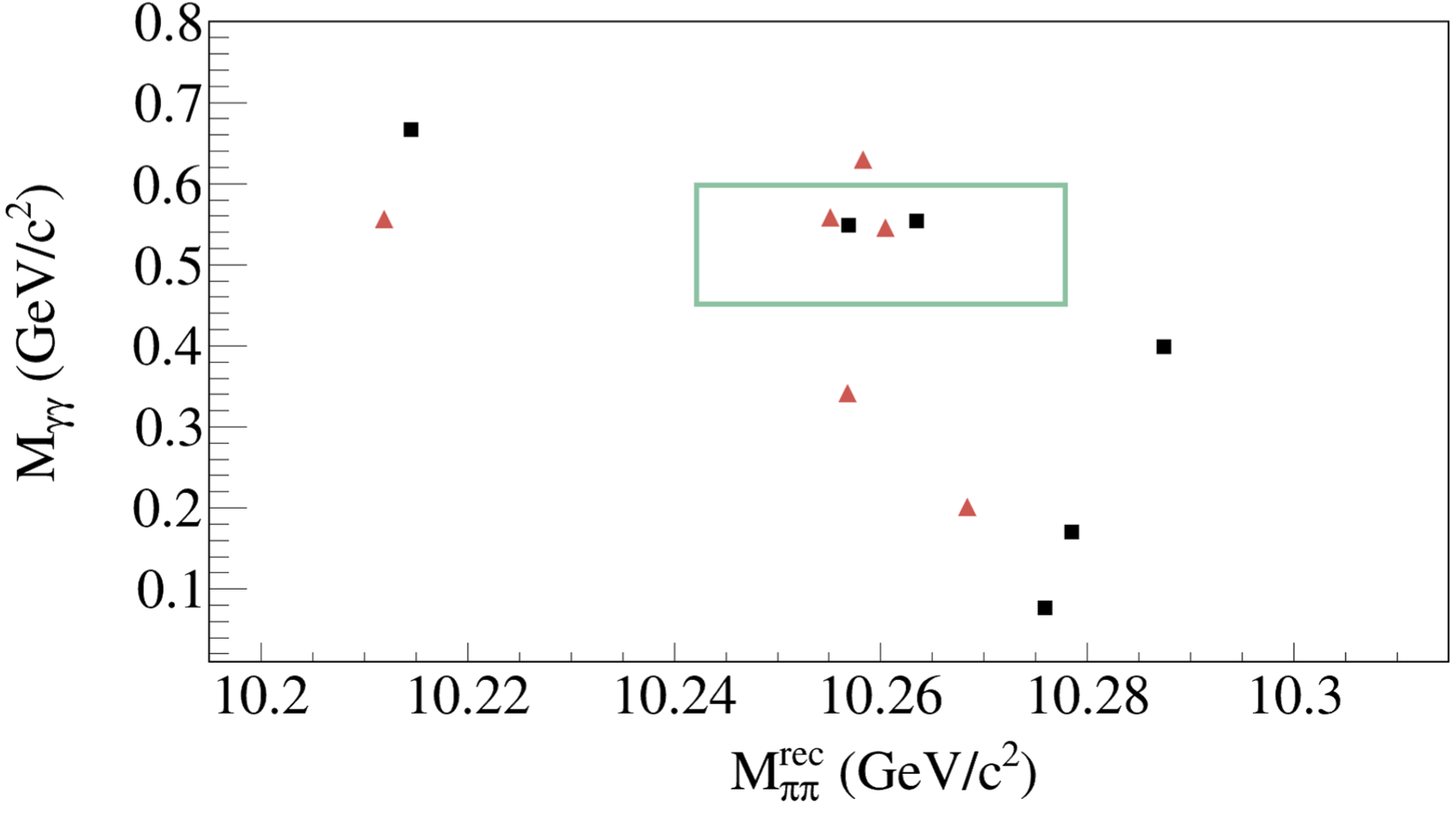}
    \caption{}
    \label{fig:U1:2d}
    \end{subfigure}
    \begin{subfigure}{0.49\linewidth}
    \includegraphics[width=\linewidth]{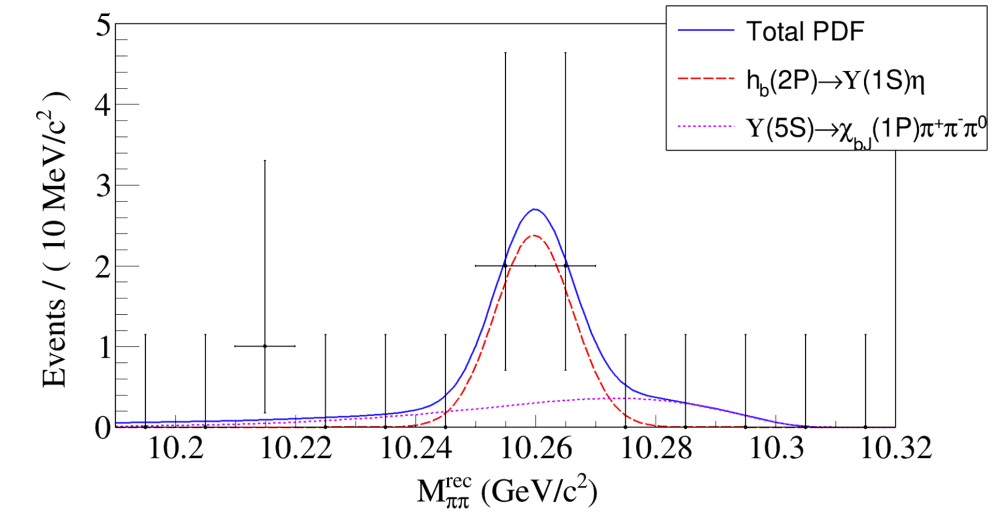}
    \caption{}
    \label{fig:U1:1d}
    \end{subfigure}
    \caption{Kinematic distributions in the process $\Upsilon(5S) \to h_b(2P) \pi^+\pi^-$ with $h_b(2P) \to \Upsilon(1S)\eta$. \subref{fig:U1:2d} shows the distribution in the invariant mass $M_{\gamma\gamma}$ of the $\gamma\gamma$ system in the decay $\eta\to\gamma\gamma$ versus the invariant mass $M_{\pi\pi}^\mathrm{rec}$ of the system recoiling against the $\pi^+\pi^-$ system, which corresponds to the $h_b(2P)$ decay products, but is calculated from the measured momenta of the $\pi^-$ and $\pi^+$ and the well-known $e^+e^-$ initial state using 4-momentum conservation. The green rectangle indicates the signal region. The red triangular markers show events where $\Upsilon(1S)\to e^+e^-$. The black square markers show events where $\Upsilon(1S)\to e^+e^-$. \subref{fig:U1:1d} shows the $M_{\pi\pi}^\mathrm{rec}$ distribution. The red curve indicates the $h_b(2P) \to\Upsilon(1S) \eta$ signal model. The violet curve indicates background from $\Upsilon(5S)\to \chi_{bJ}(1P) \pi^+\pi^-\pi^0$ decays. The blue curve indicates the total model. Taken from ref.~\cite{Belle:2024qmw}.}
\end{figure}

We study the decay $h_b(2P) \to \Upsilon(1S) \eta$ using a data sample corresponding to an integrated luminosity of \SI{121}{\ifbarn}, recorded at Belle. Details of this analysis can be found in ref.~\cite{Belle:2024qmw}.
The $h_b(2P)$ is produced in the process $e^+e^- \to h_b(2P) \pi^+\pi^-$ at a $e^+e^-$ center-of-mass energy corresponding to the $\Upsilon(5S)$ resonance. This process proceeds predominantly through intermediate $Z_b(10610)$ and $Z_b(10650)$ states in the $h_b(2P)\pi$ system.
We reconstruct the $\Upsilon(1S)$ through its decay to $\ell^+\ell^-$ and the $\eta$ through its decay to two photons.
The signal region (green rectangle in \cref{fig:U1:2d}) is defined within the two-dimensional plane of the invariant mass $M_{\gamma\gamma}$ of the $\gamma\gamma$ system corresponding to the mass of the $\eta$ and the invariant mass $M_{\pi\pi}^\mathrm{rec}$ of the system recoiling against the $\pi^+\pi^-$ system corresponding to the mass of the $h_b(2P)$. We find four events in the signal region illustrated by the black and red markers in \cref{fig:U1:2d}.
In order to extract the signal yield, an unbinned extended maximum likelihood fit is performed to the two-dimensional signal plane, taking into account the backgrounds, which arise primarily from the process $e^+e^- \to \chi_{bJ}(1P) \pi^+\pi^-\pi^0$ with $\chi_{bJ}(1P) \to \Upsilon(1S)\gamma$ and from the combinatorial background. \Cref{fig:U1:1d} shows the one-dimensional projection in $M_{\pi\pi}^\mathrm{reco}$ of the data alongside the model curves resulting from the fit.

We find evidence for the $h_b(2P) \to\Upsilon(1S)\eta$ decay with a significance of $3.5\sigma$, including systematic uncertainties, and determine that the corresponding branching fraction is $\mathcal B[h_b(2P)\to \Upsilon(1S)\eta] = (7.1^{+3.7}_{-3.2}\pm0.8)\times10^{-3}$, where the first uncertainty is statistical and the second systematic.
The measured branching fraction is approximately 10 times lower than the expected value of \SI{10}{\percent} based on the $\mathcal B[\Upsilon(3S)\to h_b(1P)\pi^0]$ decay~\cite{Li:2012is}. This discrepancy disfavors the significant contribution of loop effects to this decay process under study.

In addition, we investigate the isospin-violating decays $h_b(2P) \to \Upsilon(1S) \pi^0$ and $h_b(1P)\to \Upsilon(1S)\pi^0$ by adjusting the $M_{\gamma\gamma}$ and $M_{\pi\pi}^\mathrm{rec}$ signal regions accordingly.
We find no events in the corresponding signal regions and set upper limits for the corresponding branching fractions of $\mathcal B[h_b(2P)\to \Upsilon(1)\pi^0] < 1.8\times 10^{-3}$ and $\mathcal B[h_b(2P)\to \Upsilon(1)\pi^0] < 1.8\times 10^{-3}$ at the \SI{90}{\percent} confidence level.

\subsection{Search for $h_b(2P) \to \chi_{bJ}(1P)\gamma$}

We complement the studies of spin-singlet to spin-triplet transitions by conducting a search for the radiative $M1$ transition $h_b(2P) \to \chi_{bJ} (1P) \gamma$ with $J=0,1,2$.
In the relativistic quark model, the direct radiative $M1$ transition is strongly suppressed, with a branching fraction of $10^{-6}$ -- $10^{-5}$~\cite{Godfrey1985}. However, the branching fraction of the transition may be enhanced to $10^{-2}$ \text{--} $10^{-1}$ by couple-channel effects from hadron loops~\cite{Guo:2016yxl}.

\begin{figure}
    \centering
    \includegraphics[width=0.6\linewidth]{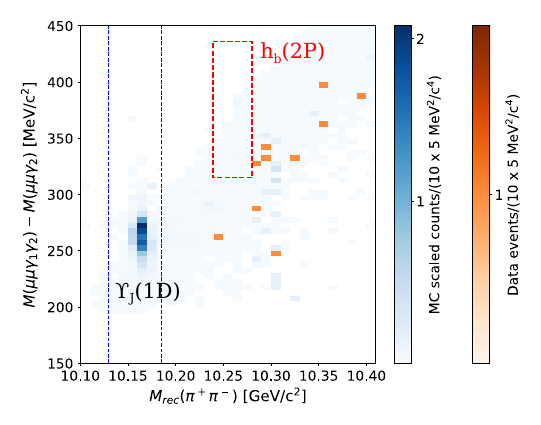}
    \caption{Kinematic distribution in the process $\Upsilon(5S) \to h_b(2P) \pi^+\pi^-$ with $h_b(2P) \to \chi_{bJ}(1P) \gamma_1$ and $\chi_{bJ}(1P) \to \Upsilon(1S)\gamma_2$. The vertical axis shows the difference of the invariant masses of the $h_b(2P)$ decay products ($M(\mu\mu\gamma_1\gamma_2)$) and the $\chi_{bJ}$ decay products ($M(\mu\mu\gamma_2)$). The horizontal axis shows the invariant mass $M_\mathrm{rec}(\pi^+\pi^-)$ of the system recoiling against the $\pi^+\pi^-$ system, which corresponds to the $h_b(2P)$ decay products, but is calculated from the measured momenta of the $\pi^-$ and $\pi^+$ and the well-known $e^+e^-$ system using 4-momentum conservation. The orange histogram shows the experimental data. The blue histogram shows the expected background. The red dashed rectangle indicates the signal region. The blue dashed lines indicate the blinded region where where $e^+e^-\to\Upsilon(5S)\to \Upsilon_J(1D)\pi^+\pi^-$ decay may be observed. Taken from ref.~\cite{Belle:2024jta}.}
    \label{fig:chi1}
\end{figure}

We study this decay using the same data sample as described in \cref{sec:U1} corresponding to an integrated luminosity of \SI{121}{\ifbarn} recorded at Belle. Details of this analysis can be found in ref.~\cite{Belle:2024jta}.
The $h_b(2P)$ is produced in the process $e^+e^- \to h_b(2P) \pi^+\pi^-$ at an $e^+e^-$ center-of-mass energy corresponding to the $\Upsilon(5S)$ resonance.
We reconstruct the $h_b(2P)$ via its decay to $\chi_{bJ}\gamma_1$, the $\chi_{bJ}(1P)$ via its decay to $\Upsilon(1S)\gamma_2$, and the $\Upsilon(1S)$ via its decay to $\mu^+\mu^-$.
Then we define the signal region (red dashed rectangle) in the two-dimensional plane shown in \cref{fig:chi1}.
The signal region in the difference of the invariant masses of the $h_b(2P)$ decay products ($M(\mu\mu\gamma_1\gamma_2)$) and the $\chi_{bJ}(1P)$ decay products ($M(\mu\mu\gamma_2)$) (vertical axis) is sufficiently large to encompass all three considered $\chi_{bJ}(1P)$ states, i.e. $\chi_{b0}(1P)$, $\chi_{b1}(1P)$, and $\chi_{b2}(1P)$.
The signal region in the invariant mass $M_\mathrm{rec}(\pi^+\pi^-)$ of the system recoiling against the $\pi^+\pi^-$ system encompasses the $h_b(2P)$ state.
In this analysis, we keep the region $10.130 < M_\mathrm{rec}(\pi^+\pi^-) < 10.185\,\si\GeVcc$ (blue dashed lines in \cref{fig:chi1}) blind, as this region is currently being studied in another analysis dedicated to search for $\Upsilon_J(1D) \to \Upsilon(1S)\gamma\gamma$ decays in the progress $e^+e^-\to \Upsilon_J(1D) \pi^+\pi^-$, i.e. in the same final state.

We identify no events (orange histogram in \cref{fig:chi1}) in the signal region and set upper limits for the corresponding branching fractions of $\mathcal{B}[h_b(2P) \to \chi_{b0}(1P)\gamma] < \num{2.7e-1}$, $\mathcal{B}[h_b(2P) \to \chi_{b1}(1P)\gamma] < \num{5.4e-3}$, and $\mathcal{B}[h_b(2P) \to \chi_{b2}(1P)\gamma] < \num{1.3e-2}$.
These upper limits are consistent with the low predictions from the relativistic quark model~\cite{Godfrey1985}. Also, they are already in the range sensitive to predictions for hadronic loop contributions~\cite{Guo:2016yxl}. However, larger data samples are required to further constrain the contributions from hadronic loops.

\section{Search for $P_{c\bar c s}(4459)^0$ and $P_{c\bar c s}(4338)^0$ in $\Upsilon(1S,2S)$ Inclusive Decays}

In addition to mesons build from a $q\bar q$ pair, such as bottomonia discussed in \cref{sec:bb}, or baryons build from $qqq$, QCD allows for complex configurations such as tetraquarks ($qq\bar q\bar q$) or pentaquarks ($qqqq\bar q$), which are called exotics.
Starting with the discovery of the $\chi_{c1}(3872)$ (a.k.a. $X(3872)$) by Belle~\cite{Belle:2003nnu}, a whole new zool of such exotic states was discovered by Belle, LHCb, and other experiments.
Currently, two candidates for pentaquarks with a $c\bar c$ pair and a strange quark ($P_{c\bar c s}$) exist, i.e. the $P_{c\bar cs}(4338)^0$ with a significance of $15\sigma$~\cite{LHCb:2022ogu} and the $P_{c\bar c s}(4459)$ with a significance of $3.1\sigma$~\cite{LHCb:2020jpq}. Both were measured by LHCb in their decay to $J/\psi \Lambda$.

We search for pentaquark states in the decays of $\Upsilon(1S)$ and $\Upsilon(2S)$ using the world's largest sample of these decays from Belle. The enhancement in the production of baryons and deuterons in $\Upsilon(1S,2S)$ decays~\cite{BaBar:2014ssg} suggests these decays to serve as a source of exotic multiquark configurations.
A search for pentaquarks without strangeness in the inclusive process $\Upsilon(1S,2S) \to P_{c\bar c} + X \to J/\psi p + X$, where $X$ indicates any further decay product, did not yield a signal~\cite{Belle:2024mcb}.

\begin{figure}
    \begin{subfigure}{0.49\linewidth}
    \includegraphics[width=\linewidth]{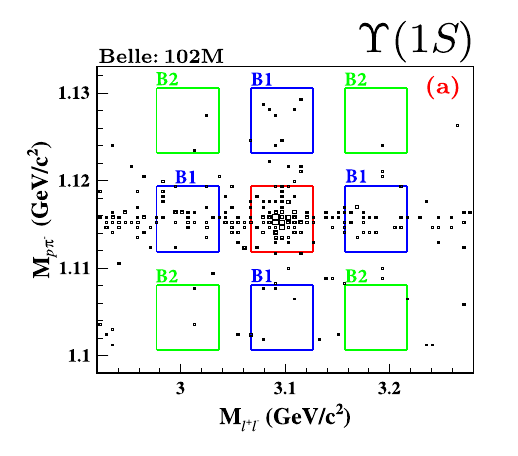}
    \caption{}
    \label{fig:Pcc:2D}
    \end{subfigure}
    \begin{subfigure}{0.49\linewidth}
    \includegraphics[width=\linewidth]{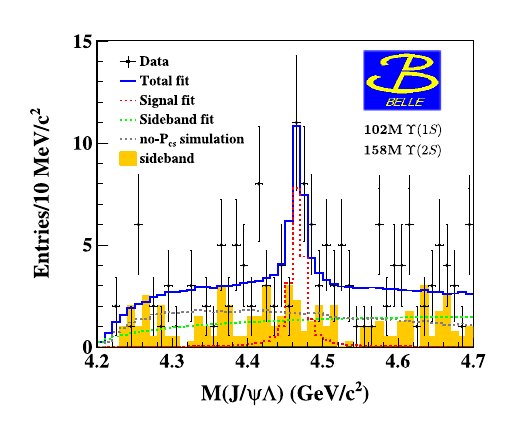}
    \caption{}
    \label{fig:Pcc:M2}
    \end{subfigure}
    \caption{Kinematic distributions in the inclusive process $\Upsilon(1S,2S) \to P_{c\bar c s} + X \to J/\psi \Lambda + X$, where the $J/\psi$ decays to $\ell^+\ell^-$ and the $\Lambda$ to $p \pi^-$. \subref{fig:Pcc:2D} shows the distribution in the invariant mass of the $\Lambda$ decay products versus the invariant mass of the $J/\psi$ decay products for $\Upsilon(1S)$ decays. The red box indicates the signal region. The blue and green boxes indicate the sideband regions. For the analogous plot for $\Upsilon(2S)$ decays see fig.~1.(b) in ref.~\cite{Belle:2025pey}. \subref{fig:Pcc:M2} shows the distribution in the invariant mass of the $J/\psi \Lambda$ system combined for $\Upsilon(1S)$ and $\Upsilon(2S)$ decays. The black markers show the real data in the signal region. The orange histogram shows the background estimated from the two-dimensional sideband regions. The solid curve indicates the total fit result. The red dashed curve indicates the signal. The brown dashed curve indicates the no-$P_{c\bar c s}$ component. The green dashed curve indicates the background from the sidebands. Taken from ref.~\cite{Belle:2025pey}.}
\end{figure}

In this analysis, we search for pentaquarks with strangeness in the inclusive process $\Upsilon(1S,2S) \to P_{c\bar c s} + X \to J/\psi \Lambda + X$, where the $J/\psi$ decays to $\ell^+\ell^-$ and the $\Lambda$ to $p \pi^-$. Details of the analysis can be found in ref.~\cite{Belle:2025pey}.
The signal region (red box) is defined in the two-dimensional plane of the invariant masses of the $\Lambda$ and $J/\psi$ decay products shown in \cref{fig:Pcc:2D}. The background in the signal region is estimated from the two-dimensional sidebands and subsequently subtracted from the signal region. The blue boxes in \cref{fig:Pcc:2D} indicate sideband regions used to determine backgrounds without $J/\psi$ or $\Lambda$ and the green boxes indicate sideband regions used determine combinatorial backgrounds.
After background subtraction, the inclusive branching fractions are $\mathcal{B}[\Upsilon(1S)\to J/\psi \Lambda/\bar\Lambda + X]= (36.9\pm5.3\pm2.4)\times10^{-6}$ and $\mathcal{B}[\Upsilon(2S)\to J/\psi \Lambda/\bar\Lambda + X] = (22.3\pm5.7\pm3.1)\times10^{-6}$.

In order to search for $P_{c\bar c s}$ pentaquarks decaying to $J /\psi /\Lambda$, we measure the signal yields in bins of the invariant mass $M(J /\psi \Lambda)$ of the $J/\psi\Lambda$ system shown as data points in \cref{fig:Pcc:M2}.
The $M(J/\psi \Lambda)$ distribution exhibits an enhancement near the nominal mass of the $P_{c\bar c s}(4459)^0$.
We perform a fit to the $M(J/\psi \Lambda)$ distribution (blue curve) to extract the $P_{c\bar c s}(4459)^0$ significance and signal yield. In addition to the $P_{c\bar c s}(4459)^0$ signal (red dashed curve), we add a background polynomial (green dashed curve). This background polynomial is further constrained by the simultaneous fitting of it to the side-band data (orange histogram). We also add a component for non-resonant, i.e. non-$P_{c\bar c s}$, contributions (brown dashed curve).
In this fit, the mass and width of the $P_{c\bar c s}(4459)^0$ are constrained to the values measured by LHCb~\cite{LHCb:2020jpq}.
From this fit, we determine the significance of the $P_{c\bar c s}(4459)^0$ signal to be $3.3\sigma$, including systematic uncertainties, and the corresponding branching fractions of $\mathcal B[\Upsilon(1S)\to P_{c\bar c s}(4459)^0 + X \to J/\psi \Lambda + X] = (3.5\pm2.0\pm0.2)\times10^{-6}$ and $\mathcal B[\Upsilon(2S)\to P_{c\bar c s}(4459)^0 + X \to J/\psi \Lambda + X] = (2.9\pm1.7\pm0.4)\times10^{-6}$.

To measure the mass and width of the $P_{c\bar c s}(4459)^0$, we perform a fit to the $M(J/\psi \Lambda)$ distribution without the mass and width constraints for LHCb. We obtained $M=4471.7\pm4.8\pm0.6\,\si\MeVcc$ and $\Gamma = 22\pm13\pm3\,\si\MeVcc$. Both are consistent with the values determined by LHCb~\cite{LHCb:2020jpq}.

Finally, we found no significant signal of the $P_{c\bar c s}(4338)^0$ in our data and set upper limits for the corresponding branching fractions of $\mathcal B[\Upsilon(1S)\to P_{c\bar c s}(4338)^0 + X \to J/\psi \Lambda + X] < \num{1.8e-6}$ and $\mathcal B[\Upsilon(2S)\to P_{c\bar c s}(4338)^0 + X\to J/\psi \Lambda + X] < \num{1.6e-6}$ at \SI{90}{\percent} confidence level.

\section{Summary}
In summary, we conduct searches for transitions from the spin-singlet $h_b(1P,2P)$ states to the spin-triplet $\Upsilon(1S)$ and $\chi_{bJ}(1P)$ states. We find no evidence for the  $h_b(1P,2P)\to \Upsilon(1S) \pi^0$ and $h_b(2P) \to \chi_{bJ}(1P)\gamma$ transitions. We find the first evidence for the $h_b(2P)\to \Upsilon(1S)\eta$ transition. However, the measured branching fraction is lower than expected from the decay of the $\Upsilon(3S) \to h_b(1P)\pi^0$ decay. Our findings disfavor the substantial contribution of couple-channel effects via hadronic loops to these transitions.
Finally, we find evidence for $P_{c\bar c s}(4459)^0 \to J/\psi \Lambda$ decays, where the $P_{c\bar c s}(4459)^0$ is produced in inclusive $\Upsilon(1S,2S)$ decays. This is the first evidence for exotic states produced in $\Upsilon(1S,2S)$ decays. It thereby lays the foundation for further searches for exotic states in these decays.
In the future, Belle~II will collect even larger data sets, thus enabling us to study hadron spectroscopy at Belle and Belle~II with unprecedented precision.


\printbibliography[heading=bibintoc]

\end{document}